\documentclass{naturefigTCL}

\usepackage{epstopdf}
\usepackage{amsmath}
\usepackage{amsbsy}
\usepackage{amsthm}
\usepackage{amssymb}
\usepackage{latexsym}
\usepackage[dvips]{color,graphicx}
\usepackage{epsfig}
\usepackage{graphicx}
\usepackage{amsfonts}
\usepackage{color}
\usepackage{ulem}

\newcommand{\llangle}{\langle\langle}
\newcommand{\rrangle}{\rangle\rangle}

\begin{document}

\title{Quantum spin-liquid emerging in two-dimensional correlated Dirac fermions}
\author{Z.~Y.~Meng$^1$, T.~C.~Lang$^2$, S.~Wessel$^1$, F.~F.~Assaad$^2$ \& A.~Muramatsu$^1$}

\maketitle

\begin{affiliations}
 \item Institut f\"ur Theoretische Physik III, Universit\"at Stuttgart, Pfaffenwaldring 57, 70550 Stuttgart, Germany
 \item Institut f\"ur Theoretische Physik und Astrophysik, Universit\"at W\"urzburg, Am Hubland, 97074 W\"urzburg, Germany
\end{affiliations}


\begin{abstract}
At sufficiently low temperatures, 
condensed-matter systems tend to develop order. An exception are quantum
spin-liquids, where fluctuations prevent a transition to an ordered state
down to the lowest temperatures. While such states are possibly realized in
two-dimensional organic compounds, they have remained elusive in
experimentally relevant microscopic two-dimensional models. 
Here, we show by means of large-scale quantum Monte Carlo simulations of 
correlated fermions on the honeycomb lattice, a structure realized in
graphene, that a quantum spin-liquid emerges between the state described by 
massless Dirac fermions and an antiferromagnetically ordered Mott insulator. 
This unexpected quantum-disordered state is found to be a short-range
resonating valence bond liquid, akin to the one proposed for high temperature
superconductors. Therefore, the possibility of unconventional superconductivity
through doping arises. We foresee its realization with ultra-cold atoms or with
honeycomb lattices made with group IV elements.
\end{abstract}


The quantum mechanical description of the relativistic electron was attained by 
Dirac, who revealed both its intrinsic angular momentum (the spin), with 
a half-integer quantum number ${S=1/2}$, and the existence of its antiparticle, the positron\cite{weinberg05}.
Both obey the Fermi-Dirac statistics, which implies that two identical particles 
cannot occupy the same quantum mechanical state. Such particles are generically called 
fermions. In case of a vanishing rest mass, the energy of Dirac fermions is a linear function of 
momentum. Such massless Dirac fermions were recently observed in two-dimensional
solid-state systems like graphene\cite{novoselov05,zhang05} and 
surfaces of bismuth based compounds\cite{zhang09,chen09}. 
For graphene, a single layer of carbon atoms with honeycomb structure, 
unusual electronic behaviour is anticipated, and partly verified experimentally, 
due to the two-dimensional Dirac-like dispersion of the electrons at low 
energies\cite{castroneto09a}. 
The interplay of a relativistic dispersion with interactions at half-filling is 
expected to lead to a quantum phase transition between the 
semimetal (SM) at low and a Mott insulator (MI) at high interaction
strengths\cite{herbut06,drut09a}. Here, a Mott insulator is an insulating state
that results not from the band structure alone, but is due to the effects of interactions.
Such correlation effects can be displayed by the Hubbard model in its most basic 
form, as exemplified in high temperature superconductivity\cite{lee06}, or with 
ultra-cold fermionic atoms loaded in optical lattices\cite{joerdens08,schneider08a}.
Studies of Hubbard-like models on the honeycomb lattice suggested the emergence 
of exotic phases such as gapless spin liquids\cite{lee05,hermele07}, charge 
density waves\cite{raghu08}, quantum spin Hall states\cite{kane05,raghu08}, or 
superconductivity\cite{uchoa07} at or near a density of one fermion per site 
(half-filling for the two-species case). 

Given the various phases proposed for fermions on a honeycomb lattice based on 
Hubbard-like models, it is important to explore the ground-state 
properties in the intermediate coupling regime of the original lattice model 
with an unbiased method. Due to the absence of a sign-problem in determinantal 
quantum Monte Carlo (QMC) simulations (see Methods) in the half-filled case, it 
is the method of choice for extrapolations to the thermodynamic limit (TDL), 
leading to essentially exact results limited only by the statistical noise.
Employing large-scale quantum Monte Carlo simulations of the spin-$\frac{1}{2}$ 
Hubbard model at half-filling on the honeycomb lattice, we show that for 
intermediate interactions a gapped non-magnetic phase destroys the semimetal 
before the transition to an antiferromagnetically ordered Mott insulator at 
strong interactions sets in. This quantum spin-liquid phase is characterized by 
local correlations that correspond to a resonating valence-bond (RVB) 
state\cite{anderson52, fazekas74} as  proposed in the context of high
temperature superconductivity\cite{anderson87,kivelson87,lee06}. 

Following their original proposals\cite{anderson52,fazekas74,anderson87,kivelson87}, 
spin-liquid states were established in effective models of singlet-dynamics such as 
quantum dimer models\cite{rokhsar88,moessner01b,moessner08}. Our results show that RVB 
states are realized in a microscopic model of correlated electrons, bringing 
closer their observation in experiments. Honeycomb lattices of group IV 
elements\cite{cahangirov09} and ultra-cold fermionic atoms loaded in optical 
lattices\cite{duan03,joerdens08,schneider08a} appear as promising candidate systems 
to realize the RVB state out of Dirac fermions.

\section*{\textbf{Phase diagram from quantum Monte Carlo}}%

Previous numerical studies of the Hubbard model on the honeycomb lattice\cite{sorella92c,paiva05}
suggested that a single quantum phase transition separates 
the paramagnetic weak-coupling SM phase from a strong-coupling 
antiferromagnetic (AF) MI. At strong enough repulsion, antiferromagnetism is 
certainly possible since the honeycomb lattice is bipartite, so that AF order 
is not geometrically frustrated. However, the honeycomb lattice has the 
smallest coordination number in two dimensions, such that the effect of quantum 
fluctuations is the strongest. Hence, the competition between the tendency to 
order and quantum fluctuations requires a detailed analysis of correlations and 
a careful extrapolation to the TDL in order to characterize the possible 
phases. Here, we present results based on projective (temperature $T=0$) 
determinantal QMC simulations in the canonical ensemble at half-filling. In 
order to assess the above scenarios, we focus in particular on the region near 
the Mott transition. 

The Hamiltonian of the spin-$\frac{1}{2}$ Hubbard model on the honeycomb lattice equals
\begin{equation}
   H=-t \sum_{\langle i,j \rangle,\alpha} ( c^\dagger_{i\alpha} c_{j\alpha} +c^\dagger_{j\alpha} c_{i\alpha} ) + U \sum_i n_{i\uparrow} n_{i\downarrow},
\end{equation}
where $c^\dagger_{i\alpha}$ ($c_{i\alpha}$) denotes the creation (annihilation) 
operator for fermions of spin $\alpha=\uparrow,\downarrow$ on lattice site $i$, 
and $n_{i\alpha}=c^\dagger_{i\alpha} c_{i\alpha}$. Here, $t$ denotes the 
nearest-neighbour hopping amplitude, and $U\geq 0$ the strength of the onsite 
repulsion. Our notations in real and momentum space are shown in the inset of 
Fig.~1. At $U=0$, the tight-binding Hamiltonian has a linear dispersion near 
the Dirac points ($K$, $K'$ -- cf.~Fig.~1), where the conduction and valence 
bands touch at half-filling, corresponding to a density $\sum_\alpha \langle n_
{i\alpha}\rangle =1$. At half-filling, the finite-$U$ region can be studied 
using projective QMC to obtain ground-state expectation values of any physical 
observable. Details are relegated to the Methods section. The phases described 
in the following correspond to extrapolations to the TDL. For that purpose we 
study lattices of $N=2L^2$ sites with periodic boundary conditions, and linear 
sizes up to $L=18$. 

To monitor the electronic properties of the system upon increasing $U$, we 
extracted the single-particle excitation gap $\Delta_{sp}(\mathbf{k})$ from the 
imaginary-time displaced Green's function (cf.~Supplementary Information (SI) 
for details). $\Delta_{sp}(\mathbf{k})$ gives the minimal energy necessary to 
extract one fermion from the system, and corresponds to the gap that can be 
observed in photoemission experiments. As shown in Fig.~1, $\Delta_{sp}(K) = 0$ 
for $U/t$ below about 3.5, as expected for a SM. For larger $U/t$, the system 
enters into an insulating phase due to interactions. The values of the gap are 
obtained via an extrapolation of the QMC data to the TDL as shown in Fig.~2a.

From previous analysis of the model, one expects long-range AF correlations 
when the MI appears. We therefore measured the AF spin structure factor $S_{AF}
$ (cf. SI) that reveals long-range AF order if $m_s^2=\lim_{N \rightarrow 
\infty} S_{AF}/N >0$. Figure 2b shows the QMC results together with a finite 
size extrapolation. The results of the latter are also presented in the phase 
diagram of Fig.~1. AF order appears beyond $U/t=4.3$, a value that is 
consistent with previous estimates for the onset of long-ranged AF 
order\cite{sorella92c,paiva05}. This leaves an extended window 
$3.5< U/t <4.3$, within which the system is neither a SM, nor an AF MI. 

Further details on the nature of this intermediate region are obtained by 
examining the spin excitation gap, extracted from the long-time behaviour of 
the imaginary-time displaced spin-spin correlation function (cf. SI). We 
consider first the spin gap $\Delta_{s}$ in the staggered sector at $\mathbf{k}=
\Gamma$, which vanishes inside the AF phase due to the emergence of two 
Goldstone modes, as well as in the gapless SM phase. Figure 2c shows finite 
size estimates of $\Delta_{s}$ for different values of $U/t$, along with an 
extrapolation to the TDL. A finite value of $\Delta_{s}$ persists within an 
intermediate parameter regime $3.5<U/t <4.3$, while it vanishes both within the 
SM and the AF phase. This dome in the spin gap is also seen in the inset of 
Fig.~2c, that displays both the finite-size data and the extrapolated values 
of $\Delta_{s}$ as a function of $U/t$. We also calculated the uniform spin gap 
$\Delta_{u}$ by extrapolating the spin gap observed at the smallest finite $
\mathbf{k}$-vector on each cluster to the TDL. $\Delta_{u}$ is found to be even 
larger than $\Delta_{s}$ inside the intermediate region (e.g. $\Delta_{u}
=0.099\pm 0.001$ (s.e.m.) at $U/t=4$), and vanishes in the SM and the 
AF phase ($\Delta_{u}$ cannot be measured directly at $\mathbf{k}=0$, because the 
uniform magnetization is a conserved quantity, cf. SI). 
Hence, this intermediate insulating region corresponds to a spin-gap phase. 

From analysing the $U$-dependence of the kinetic energy density, $E_{kin}=\langle -t 
\sum_{\langle i,j \rangle,\alpha} ( c^\dagger_{i\alpha} c_{j\alpha} +c^\dagger_
{j\alpha} c_{i\alpha} )\rangle/N$, we obtain further insight into these 
different regimes and the emergence of local moments. As shown in Fig.~3, the 
curvature $d^2E_{kin}/dU^2$ changes sign near $U/t=4.3$. This marks a 
characteristic change from the weak-coupling region of positive curvature with 
delocalized electrons to the strong-coupling AF region with negative curvature. 
In the latter region, localized spins form and order in an AF state. In the 
intermediate spin-gap region, fluctuations are large enough to still prevent 
the formation of well-localized magnetic moments. Note, that around $U/t=3.5$, 
a change in the curvature can be observed, that adds to the already presented 
evidence for an intermediate phase. 

\section*{\textbf{Characterization of the spin-gap phase}}%

The observation of a finite spin gap rules out a gapless spin-liquid\cite{lee05,hermele07},
quantum spin Hall states\cite{raghu08}, as well as triplet 
superconductivity\cite{uchoa07}. The remaining possibilities can be enumerated 
by considering the coupling to order parameters that lead to the opening of a 
mass gap in Dirac fermions\cite{Ryu09}, and hence account for the single-particle
gap observed in the QMC data: (i) singlet superconductivity, (ii) a 
quantum Hall state (QHS)\cite{Haldane98}, (iii) charge density wave (CDW) order\cite{raghu08}
and (iv) a valence bond crystal (VBC). 

In order to assess if superconductivity arises in the vicinity of the Mott 
transition, we used the method of flux quantisation which probes the superfluid 
density and is hence independent of the specific symmetry of the pair wave 
function\cite{assaad93}. Let $\Phi$ be a magnetic flux traversing the centre of 
a torus on which the electronic system lies and $E_0(\Phi/\Phi_0)$ the total 
ground-state energy, $\Phi_0$ being the flux quantum. A superconducting state 
of Cooper pairs is present if in the TDL, the macroscopic energy difference $
{E_0(\Phi/\Phi_0) - E_0(\Phi/\Phi_0 = 1/2)}$ is a function with period $1/2$\cite{byers61}.
In contrast, a metallic phase is characterized by ${E_0(\Phi/\Phi_0) - E_0(\Phi/\Phi_0 = 1/2)}$ 
vanishing as a power law as a function of system size, while in an insulating phase, 
it would vanish exponentially. As shown in the SI, this quantity vanishes in the 
TDL both in the semi-metallic state at $U=0$ and at $U/t=4$, i.e. in the 
intermediate phase. In addition, we measured pair correlations, ruling out 
superconductivity in (extended) $s$-, $p$-, $d$-, and $f$-wave channels (cf.~SI). 
Hence, both flux quantization as well as a direct measurement of pair correlations 
lead to no sign of superconductivity.

Both the CDW and QHS trigger a breaking of the sub-lattice symmetry and thereby 
open a mass gap at the mean-field level. A detailed analysis of the charge-charge 
correlation functions rules out a CDW. Furthermore, we find no signature for the 
presence of (spin) currents in the ground-state (cf.~SI). This rules out the 
breaking of sublattice and time reversal symmetries, as required for the QHS, 
in the pristine Hubbard model, and possibly, extensions of it are necessary to 
reach such a state\cite{raghu08}.

To examine the occurrence of a VBC, we probed for dimer-dimer correlations 
between separated dimers formed by nearest neighbour bonds $\langle ij\rangle$ 
and $\langle kl\rangle$ (cf.~SI). We find no VBC, neither in the charge, nor 
in the spin sector. Figure 4 shows the results of this measurement in the spin 
sector, i.e.\ the correlation between singlet dimers at ${U/t=4.0}$. The 
striped bond is the one with respect to which correlations were determined. 
They are found to be short-ranged, and consistent with the dominance of a 
RVB state within the hexagons of the honeycomb 
lattice. This can be seen by comparing the singlet-correlations with those of an 
isolated hexagon (inset Fig.~4), the classical example of the resonance 
phenomenon in conjugated $\pi$-electrons\cite{pauling39}. Accordingly, we find 
no long-ranged order from the dimer-dimer structure factors in Fourier space. 
Our results thus reveal a genuinely exotic state of matter, where no spontaneous 
symmetry breaking is observed,  while a spin-gap is present. It corresponds to a 
spin-liquid RVB state in the intermediate coupling regime in the vicinity of the 
Mott transition. 

\section*{\textbf{Further insight into the RVB state}}%

The QMC results presented above uncover the realization of a quantum spin-liquid 
state of correlated fermions on a non-frustrated, bipartite lattice. In principle, 
such quantum-disordered states can occur in different flavors, and we thus aim to 
shed further light on the nature of the spin-liquid observed above. 
Gapless (algebraic) spin-liquids, or long-range RVB states are characterized by 
critical spin-spin correlations\cite{Rantner02,Hermele04,Assaad04,lee06,imada06}.
The observation of a finite spin-gap clearly excludes such candidate states, 
while being consistent with the characterization of the observed spin-liquid in 
terms of short-range RVB states\cite{anderson52,fazekas74,kivelson87,rokhsar88,moessner01b},
also in accordance with the observed short-ranged dimer-dimer correlations.
Short-range RVB states are modeled in general by quantum dimer 
models\cite{rokhsar88,moessner01b,moessner08}, which capture the
fluctuations of singlets in a RVB state, with dimers being a
strong-coupling representation of nearest-neighbour singlets\cite{rokhsar88}.
Depending on the lattice geometry, quantum dimer models 
can exhibit a fully gapped short-range RVB phase, as e.g. on the triangular 
lattice\cite{moessner01a}, but also exhibit spin liquid states with gapless 
excitations, as is the case for bipartite lattices at the Rokhsar-Kivelson
point\cite{rokhsar88} and within the U(1) spin-liquid phase stabilized for 
spatial dimensions $d>2$\cite{fradkin04}. In the later case, gapless singlet 
excitations constitute an emerging ``photon'' soft mode.
Fully gapped phases of quantum dimer models are furthermore characterized by a 
non-trivial topological order, implying e.g.\ an emerging ground-state 
degeneracy of two-dimensional systems with periodic boundary conditions
in the TDL\cite{wen91,moessner08}.

In order to assess, whether topological order can characterize the short-range 
RVB in our case, we examine the low-energy singlet excitations. As proven in 
an exact theorem by Lieb\cite{lieb89}, the finite systems used in our 
numerical simulations have a non-degenerate singlet ground-state for any 
finite value of $U>0$. Hence, degeneracy can only appear in the TDL. If so, 
low-energy singlet states should be present, with decreasing excitation energy 
as the system size increases. Since our QMC method projects out the finite 
system's ground-state from a singlet trial wavefunction, we can monitor the 
expectation value of the internal energy $E (\Theta)$, where $\Theta$ is the 
projection parameter (cf. Methods). Here all contributions from singlet states 
with the same quantum numbers as the ground-state are included, that have a 
finite overlap with the trial wave function. For a given system of size $N$, 
we define $\Theta^*$ as the value of the projection parameter, such that 
$(E(\Theta) - E_0)/N < \epsilon$, for $\Theta > \Theta^*$, where $E_0$ is the 
ground-state energy, and $\epsilon$ is an energy scale which we choose small 
enough, in order to guarantee that $\epsilon \ll (E_1 - E_0)/N$, where $E_1$ 
is the lowest singlet excited state above the ground-state. Typically, we 
choose $\epsilon$  of the order of our statistical error in the energy density. 
Then, $1/\Theta^*$ is a lower bound for the lowest singlet excitation (cf. SI).
Such an analysis on various lattice sizes leads to the conclusion, that the 
singlet excitation gap stays above the spin gap  in the TDL (cf. SI), providing 
no evidence for the emergence of a topological state. However, we cannot 
definitely exclude topological order, if the relevant singlet states happen to 
have a vanishing overlap with our trial wave function. For the future, it will 
be interesting to explore the low-energy singlets beyond the projective scheme, 
and probe for soft modes similarly as does e.g. the construction of finite 
momentum trial states in quantum dimer models~\cite{rokhsar88,moessner08}.
Our findings, based on a controlled numerical framework, therefore open a new 
facet of quantum spin-liquids, where an appreciable amount of doubly-occupied 
sites are present, extending well beyond the regime of localized spin physics.

\section*{\textbf{Discussion and outlook}}%

\begin{sloppypar}%
The presence of a spin-liquid in the Hubbard model on the bipartite honeycomb 
lattice close to an antiferromagnetic Mott insulator resembles the situation in 
the organic antiferromagnet ${\kappa\mbox{-(BEDT-TTF)}_2\mbox{Cu}_2\mbox{(CN)}_3}$, 
which has been argued to display a spin-liquid state\cite{shimizu03,yamashita08},
albeit the latter system is on a triangular lattice and hence 
frustrated. This difference can be reconciled starting from the strong-coupling 
limit of the Hubbard model, i.e.\ a nearest-neighbour Heisenberg model, that 
close to the Mott transition acquires corrections that induce  efficient 
frustrations to the spin degrees of freedom. In fact, a Klein Hamiltonian for a 
spin-liquid state on the honeycomb lattice was constructed, including extended 
exchange interactions\cite{chayes89}. A more pronounced difference is the 
appearance of superconductivity in the organic systems upon pressure, that is 
equivalent to a reduction of the ratio $U/t$ in the Hubbard model\cite{nam07}. 
The absence of superconductivity in our system could be due to the vanishing 
density of states at the Fermi energy. In this case, a finite coupling strength 
is needed, at least in the BCS-frame\cite{kopnin08}. 
However, having an unexpected realisation of a short-range RVB state, it would 
be highly interesting to explore the consequences of doping, in a spirit rather 
close to the original scenario proposed by Anderson\cite{anderson87} and 
Kivelson \textit{et al.}\cite{kivelson87} for the cuprates.
In particular, for the fully gapped short-range RVB state, the finite spin-gap 
sets the energy scale of pairing in the superconducting state\cite{kivelson87}.
In this respect, the value obtained for the spin-gap is rather promising. The 
largest value attained is $\Delta_s \sim 0.025 t$ (Fig.~1), that for $t$ in the 
range of 1.5 to 2.5 eV (in graphene is $t=2.8$ eV\cite{castroneto09a}) 
corresponds to a temperature scale ranging from 400 to 700 K. 
\end{sloppypar}%

Although studies of doping are beyond the power of our quantum Monte Carlo 
approach due to the sign problem, they could open 
interesting perspectives e.g.\ in future experiments with ultra-cold atoms on
a honeycomb optical lattice, or with honeycomb lattices based on group IV
elements like expanded graphene (to enhance the ratio $U/t$) or Si, where the 
nearest neighbour distance is expected to be approximately 50\% larger than 
in graphene\cite{cahangirov09}, such that correlations effects are enhanced.
In fact, first attempts succeeded in synthesizing single-crystal silicon
monolayers\cite{nakano06}. 


\begin{methodssummary}

At half-filling, the finite-$U$ region can be studied using the 
determinantal projective QMC algorithm to obtain ground-state expectation values
of a physical observable by performing an imaginary time evolution of a trial wave
function that is required to be nonorthogonal to the ground-state. The value $\Theta$ 
reached in the imaginary time evolution corresponds to a projection parameter\cite{koonin68,Furukawa91,Assaad08}.
For a spin-singlet trial wave function, we found ${\Theta = 40/t}$ to be
sufficient to obtain converged ground-state quantities within statistical
uncertainty. In the presented simulations, we used a finite 
imaginary time step ${\Delta\tau = 0.05/t}$. We verified by extrapolating $
{\Delta\tau \rightarrow 0}$ that this finite imaginary time step produces no 
artifacts. Simulations were performed for systems of linear size 
$L=3,6,9,12,15$ and $18$ with $N=2L^{2}$ sites. For periodic boundary 
conditions these clusters all have nodal $K$-points and hence allow a smooth 
extrapolation to the TDL. Imaginary time displaced quantities are obtained by
using the approach in Ref.~50.

\end{methodssummary}



\begin{addendum}
 \item[Supplementary Information] is attached after the section \textbf{Methods}.
 \item[Acknowledgements] We thank L.~Balents, S.~Capponi, A.~H.~Castro Neto, 
A.~Georges, M. Hermele, A.~L\"auchli,  E.~Molinari, Y.~Motome, S.~Sachdev, 
K.~P.~Schmidt and S.~Sorella for discussions. We are grateful to S.~A. Kivelson 
for thoroughly reading our manuscript and providing important suggestions. 
F.F.A. is grateful to the KITP Santa Barbara for hospitality and acknowledges 
support by the DFG through AS120/4 and FG1162. A.M. thanks the Aspen Center for 
Physics for hospitality and acknowledges partial support by the DFG through 
SFB/TRR21. S. W. acknowledges support by the DFG through SFB/TRR21 and WE3649. 
We thank NIC J\"ulich, HRL Stuttgart, the BW Grid and the LRZ M\"unchen for the 
allocation of CPU time.
\end{addendum}
\vfill

\begin{figure}
   \centering
   \includegraphics[width=0.65\textwidth]{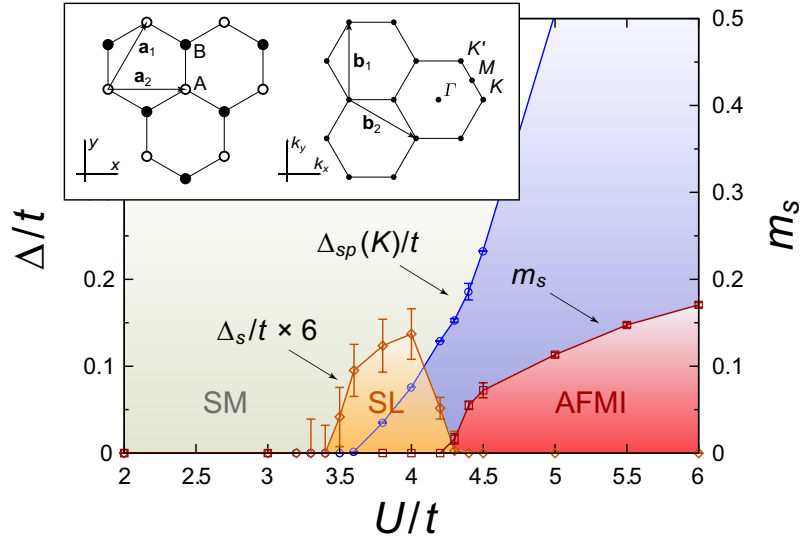}
   \caption{
\textbf{Phase diagram for the Hubbard model on the honeycomb lattice at half-filling}.
The semimetal (SM) and the antiferromagnetic Mott insulator 
(AFMI) are separated by a gapped spin liquid 
(SL) phase in an intermediate coupling regime. $\Delta_{sp}(K)$ denotes the 
single-particle gap and $\Delta_s$ the spin gap. $m_s$ denotes the staggered 
magnetization whose saturation value is $1/2$. Error bars indicate the standard 
error (s.e.m.). The inset shows the honeycomb lattice with primitive vectors $\mathbf
{a}_{1}$, $\mathbf{a}_{2}$, and reciprocal lattice vectors $\mathbf{b}_{1}$, $
\mathbf{b}_{2}$. Open (full) circles for sublattice $A$ ($B$), the Dirac points $K
$ and $K'$, and the $M$ and $\Gamma$ point are indicated.
}
\end{figure}

\begin{figure}
   \centering
   \includegraphics[width=0.42\textwidth]{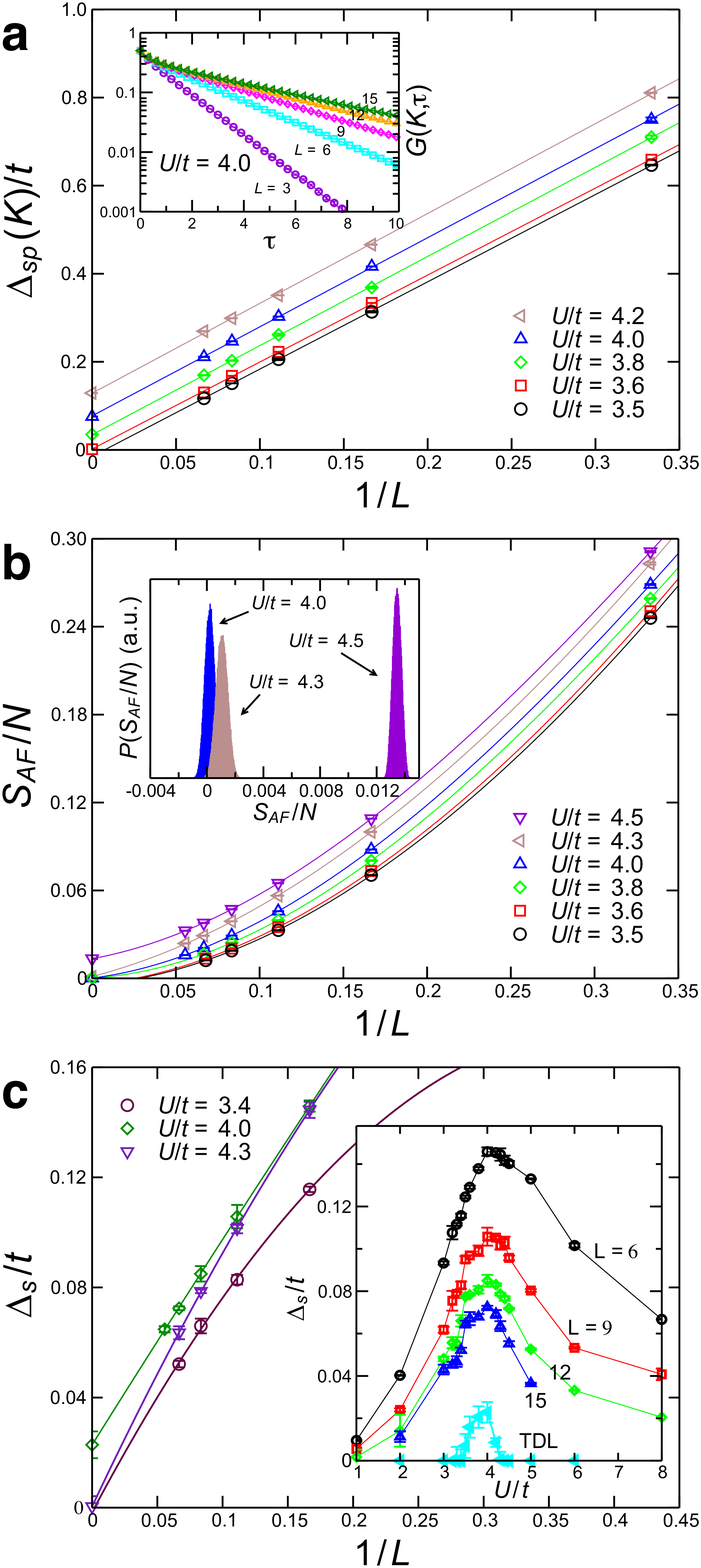}
   \caption{\textbf{Finite size extrapolations of the excitation gaps and the 
antiferromagnetic structure factor}. \textbf{a}, Single-particle gap at the 
Dirac point $\Delta_{sp}(K)$ for different values of $U/t$, linear in $1/L$. 
$\Delta_{sp}(K)$ is obtained by fitting 
the tail of the Green's function (inset) to the form $ e^{-\tau \Delta_{sp}
(K) } $. \textbf{b}, Antiferromagnetic structure factor $S_{AF}$ for various 
values of $U/t$ using $3$rd order polynomials in $1/L$. AF order appears beyond 
$U/t =4.3$, as seen in the histograms from a Monte Carlo bootstrapping analysis 
(inset). \textbf{c}, Spin gap $\Delta_{s}$ at different values of $U/t$, using 
$2$nd order polynomials in $1/L$. Error bars in \textbf{a}, \textbf{b}, and 
\textbf{c} indicate the standard error (s.e.m.).}
\end{figure}

\begin{figure}
   \centering
   \includegraphics[width=0.6\textwidth]{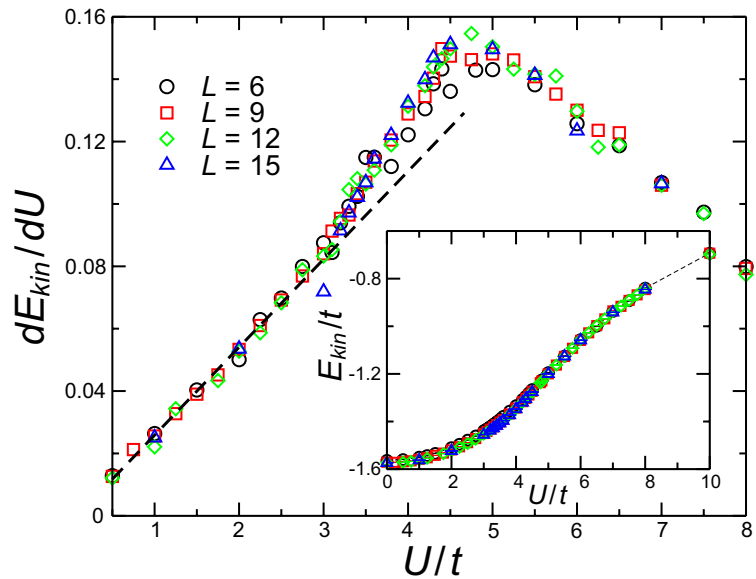}
   \caption{\textbf{Derivative $dE_{kin}/dU$ of the kinetic energy density as a 
function of $U/t$ for systems of different sizes}. The dashed line is a fit to 
the low-$U$ behaviour. The inset shows the QMC data for the kinetic energy density $E_
{kin}$ from which the derivative is obtained by numerical differentiation. 
Statistical errors (s.e.m.) are smaller than the symbol size.}
\end{figure}

\begin{figure}
   \centering
   \includegraphics[width=0.6\textwidth]{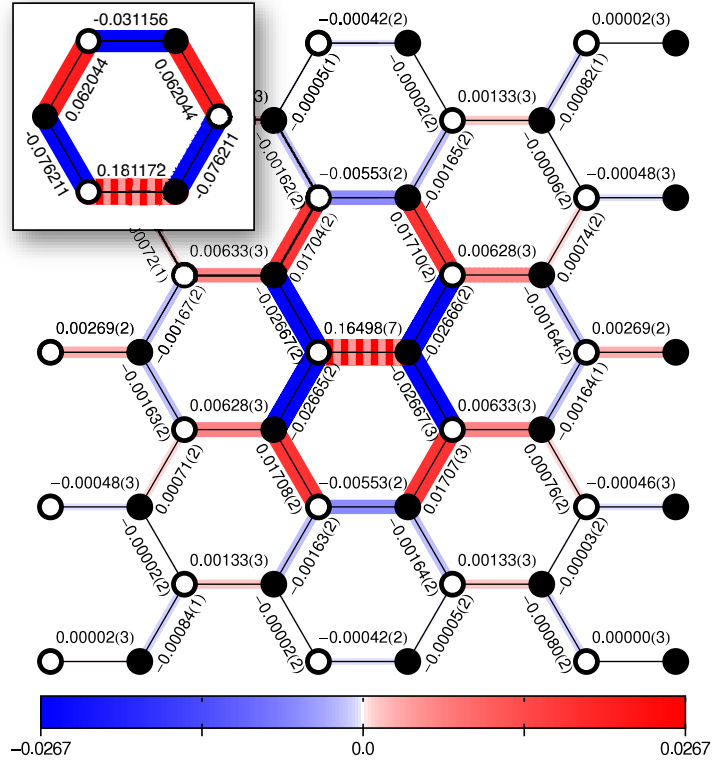}
   \caption{\textbf{Real space plot of the spin dimer-dimer correlations}. The 
correlation function $D_{ij,kl}$ (cf. SI) for a $L=6$ system at $U/t = 4$, 
together with the same correlation for the isolated Hubbard hexagon also at 
$U/t = 4$ (inset). The reference bonds are dressed with stripes. Numbers in 
parenthesis indicate the standard error (s.e.m.) of the last digit.}
\end{figure}
\vfill


\newpage
\begin{methods}

The projective QMC algorithm employed for the simulations presented in this 
article constitutes an unbiased, controlled and numerically exact method which 
is described in detail in Refs.~\arabic{ref_koonin68},\arabic{ref_Assaad08}. 
Within this scheme, ground-state expectation values of a physical observable 
$A$ are obtained from performing an imaginary time evolution
\begin{equation*}
   \langle A \rangle =\lim_{\Theta\rightarrow \infty} {\langle \Psi_T | \ensuremath{\mathrm{e}}^{-\Theta H/2} A \ensuremath{\mathrm{e}}^{-\Theta H/2} | \Psi_T\rangle } /{\langle \Psi_T|\ensuremath{\mathrm{e}}^{-\Theta H} | \Psi_T\rangle}.
\end{equation*}
Here, we used the fact that the ground-state of the Hubbard model on the 
honeycomb lattice is non-degenerate on any finite lattice at half-filling 
and for an equal number of sites within the two sublattices\cite{lieb89}, 
and implicitly assumed that  the trial wave function, $|\Psi_T\rangle$, 
has a finite overlap with this ground-state. The standard implementation of 
the algorithm  requires the trial wave function to  be a single Slater 
determinant. 

The efficiency of the projective approach strongly depends on the choice 
of the trial wave function.  To generate optimal trial wave functions, 
different approaches can be employed. One possible strategy consists in 
optimizing the overlap of the trial wave function with the finite system's 
ground-state\cite{Furukawa91}. Alternatively, one can specify a series of 
good quantum numbers that characterise the ground-state. The trial wave 
function is then constructed as to share the same quantum numbers. We have 
chosen the latter approach for our simulations, and generated 
the trial wave function from the non-interacting  tight binding model on a 
torus, through which we thrust a magnetic flux $\Phi$, corresponding to 
the vector potential ${\mathbf{A} = \Phi \frac{\mathbf{a}_1}{L}}$. In particular, 
we employed a trial wave function of the form ${|\Psi_T\rangle = |\Psi_T\rangle_
{\uparrow} \otimes |\Psi_T\rangle_{\downarrow}}$, where $|\Psi_T\rangle_\alpha$ 
denotes the ground-state of the single particle Hamiltonian in the spin-flavor 
$\alpha$ Hilbert subspace,
\begin{equation*}
	H^0_\alpha  =  -t \sum_{\langle i,j\rangle} c^{\dagger}_{i,\alpha}  c_{j,\alpha} \exp\left(\frac{2\pi\ensuremath{\mathrm{i}}}{\Phi_0}\int_i^j \mathrm{d} {\boldsymbol\ell}\cdot \mathbf{A}\right)
	+ \mathrm{H.c.} \;.
\end{equation*}
Here, ${\Phi_{0} = h e/c}$ denotes the flux quantum. At $\Phi = 0$ and for 
the considered finite lattices of linear size $L=3n$ ($n \in \mathbb{N}$),
the half-filled ground-state wave function of the above Hamiltonian 
$H^0_\alpha$ is degenerate.  Imposing an infinitesimal twist
(we verified that taking ${\Phi/\Phi_0 = 0.0001}$ is  sufficiently weak) 
lifts the two-fold degeneracy of the single particle states at the Dirac 
points $K$ and $K'$. The thereby produced filled shell configuration 
guarantees the absence of a negative sign problem, and has total momentum 
${K + K' = G}$ ($G$ being a reciprocal lattice vector) and zero total spin.
We used this trial wave function for our zero-flux quantum Monte Carlo 
simulations. Employing this trial wave function, we found an imaginary time 
projection parameter ${\Theta = 40/t}$ to be  sufficient to obtain converged 
ground-state quantities within the statistical uncertainty. 

For the presented simulations, we used an SU(2) symmetric, discrete 
Hubbard-Stratonovich transformation which allows for a direct generalization 
of the simulation scheme to SU(N) symmetric models\cite{Assaad04}. In this 
approach, after performing the standard Trotter-Suzuki 
decomposition~\cite{Assaad08}, the interaction part of the imaginary time 
evolution operator is expressed as 
\begin{equation*}
   \ensuremath{\mathrm{e}}^{-\Delta\tau U (n_{\uparrow}+n_{\downarrow} - 1 )^2/2} = 
   \sum_{l=\pm 1,\pm 2} \gamma(l)
   \ensuremath{\mathrm{e}}^{\ensuremath{\mathrm{i}}\sqrt{\Delta\tau U/2}\, \eta(l) (n_{\uparrow}+ n_{\downarrow} - 1) } + \mathcal{O}(\Delta\tau^{4}) \;,
\end{equation*} with the two functions $\gamma$ and $\eta$ of the four-valued 
auxiliary field $l=\pm 1,\pm 2$ taking on the values
\begin{eqnarray*}
   \gamma(\pm 1) = 1+\sqrt{6}/3 \;,&\quad& \eta(\pm 1) = \pm\sqrt{2\,(3-\sqrt{6})} \;,\nonumber\\
   \gamma(\pm 2) = 1-\sqrt{6}/3 \;,&\quad& \eta(\pm 2) = \pm\sqrt{2\,(3+\sqrt{6})} \;.
\end{eqnarray*}
The advantage of this representation is the fact, that for each 
Hubbard-Stratonovich configuration, the SU(2) spin symmetry of the Hubbard 
model is conserved explicitly. The above Hubbard-Stratonovich
transformation produces an overall systematic error proportional to $\Delta
\tau^3$ in the Monte Carlo estimate of  observables which,  in comparison 
to the Trotter error of order $\Delta \tau^2$, is however negligible. We 
employed a finite imaginary time step ${\Delta\tau = 0.05/t}$ and 
verified  upon extrapolating ${\Delta\tau \rightarrow 0}$, that this 
value produces no artifacts. In order to extract the gaps to the various 
excitations of the system, we calculated in addition to equal-time 
correlations also  imaginary time displaced correlation functions.
To efficiently calculate these imaginary time displaced quantities, we 
used an  approach that was introduced in Ref.~\arabic{ref_Feldbacher01}, 
and which accounts for the fact, that for a given auxiliary field 
configuration the equal-time Green-function matrix is a projector.

Finally, we have confirmed the validity of our implementation against 
exact diagonalization results on both $L=2$ and $L=3$ lattices. 

\end{methods}


\newpage

\begin{SI}
\renewcommand\figurename{\sffamily\noindent Supplementary Figure}
\setcounter{figure}{0}

\noindent In these supplementary sections, we provide further details about the simulation results mentioned in the main text. As a convenient notation, in the following  $c_{\mathbf{x}A\alpha}^{\dagger}$ and $c_{\mathbf{x}B\alpha}^{\dagger}$ ($c_{\mathbf{x}A\alpha}$ and $c_{\mathbf{x}B\alpha}$)  denote creation (annihilation) operators for fermions of spin $\alpha = \uparrow$ or $\downarrow$, on the lattice site that belongs to the sublattice $A$ and $B$ respectively, within the unit cell at position $\mathbf{x}$. Furthermore, $n_{\mathbf{x}a\alpha}=c^\dagger_{\mathbf{x}a\alpha} c^{}_{\mathbf{x}a\alpha}$ and $n_{\mathbf{x}a}=\sum_\alpha n_{\mathbf{x}a\alpha}$ denote the local density operators, and $\mathbf{S}_{\mathbf{x}a}=\frac{1}{2} c^\dagger_{\mathbf{x}a\alpha}{\boldsymbol\sigma}_{\alpha\beta}c^{}_{\mathbf{x}a\beta}$ the local spin operators, where ${\boldsymbol\sigma}=(\sigma_x,\sigma_y,\sigma_z)$ is the vector of Pauli matrices and $a\in\{A,B\}$. The corresponding operators in momentum space are obtained from
\begin{equation}
   c_{\mathbf{k}a\alpha}=\frac{1}{L^2}\sum_{\mathbf{x}} e^{-i\mathbf{k}(\mathbf{x}+\mathbf{x}_a)} c_{\mathbf{x}a\alpha},
\end{equation} 
where $\mathbf{x}_A=(0,0)$ and  $\mathbf{x}_B=(0,a)$, with $a$ the distance between neighbouring lattice sites. Similarly, Fourier components $n_{\mathbf{k}a\alpha}$, $n_{\mathbf{k}a}$ and $\mathbf{S}_{\mathbf{k}a}$ of the density and spin operators are defined. For the following, it is  also convenient to introduce the three lattice vectors related to the three nearest neighbour bonds,
\[
{\boldsymbol\delta}_{1}=(0,0), \quad {\boldsymbol\delta}_{2}=-\mathbf{a}_2, \quad {\boldsymbol\delta}_{3}=\mathbf{a}_1-\mathbf{a}_2,
\]
where $\mathbf{a}_1$ and $\mathbf{a}_2$ are shown in Fig.\ 1, as well as
\[
\mathbf{r}_1=\mathbf{a}_2,\quad \mathbf{r}_2=\mathbf{a}_2-\mathbf{a}_1,  \quad \mathbf{r}_3=-\mathbf{a}_1,\quad \mathbf{r}_4=-\mathbf{r}_1,\quad \mathbf{r}_5=-\mathbf{r}_2,\quad \mathbf{r}_6=-\mathbf{r}_3
\] 
connecting a given lattice site to its six next-nearest neighouring lattice sites. 
For the correlation between two local operators $O_1$ and $O_2$, we employ a short notation for the cummulant, 
\begin{equation}
   \llangle O_1 O_2\rrangle := \langle O_1 O_2\rangle - \langle O_1\rangle \langle O_2\rangle.
\end{equation}
Most of the following results concern the intermediate spin liquid phase, and we present in those cases quantum Monte Carlo data for the representative value of $U/t=4$. 

\section{Green's function and single-particle gap}
To probe the single-particle properties, we measured the imaginary-time displaced Green's function
\begin{equation}
G(\mathbf{k},\tau)= \frac{1}{2}
\sum_{a} {\langle c^\dagger_{\mathbf{k}a\uparrow}(\tau) c_{\mathbf{k}a\uparrow}(0) \rangle }=
\frac{1}{2}
\sum_{a} {\langle c^\dagger_{\mathbf{k}a\downarrow}(\tau) c_{\mathbf{k}a\downarrow}(0) \rangle },
\end{equation} 
where $c^{(\dagger)}_{\mathbf{k}a\alpha}(\tau)=e^{\tau H} c^{(\dagger)}_{\mathbf{k}a\alpha}e^{-\tau H}$. 
The single-particle gap $\Delta_{sp}(\mathbf{k})$ is obtained from $G(\mathbf{k},\tau) \propto \exp(-\tau\Delta_{sp}(\mathbf{k}))$ at large imaginary time $\tau$,
and corresponds to the particle (or hole) excitation energy with respect to the chemical potential $\mu=0$ at half-filling in this particle-hole symmetric system.
At $U=0$, the single-particle gap vanishes at the Dirac points $K$ and $K'$ (cf. Fig.\ 1 for our notation in momentum space), and we thus considered $\Delta_{sp}(K)$ in detail. The quantum Monte Carlo data for $G(K,\tau)$ and $\Delta_{sp}(K)$ is presented in the main text. 

\section{Spin correlations and $S_{AF}$}
The antiferromagnetic order at large values of $U/t$ resides within the unit cell of the honeycomb lattice. 
Hence, the spin structure factor for antiferromagnetic order relates to the staggered spin correlations at the $\Gamma$ point (cf. Fig.\ 1 for our notation in momentum space),
\begin{equation}
S_{AF}=\langle  [\sum_{\mathbf{x}}(\mathbf{S}_{\mathbf{x}A}-\mathbf{S}_{\mathbf{x}B})]^2/N\rangle.
\end{equation}
\begin{figure}[hb]
\centering
\includegraphics[width=8cm]{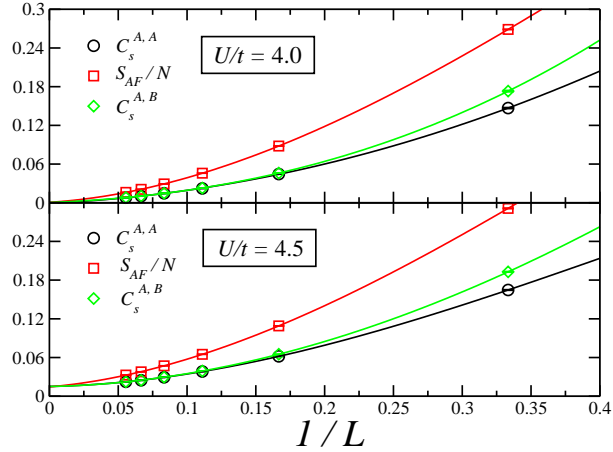}
\caption{Comparison of the finite size scaling between the spin correlations $C^{A,A}_{s}$ and $C^{A,B}_{s}$ at the largest available distance  and the staggered structure factor $S_{AF}$ at $U/t = 4$ (upper panel) and $U/t=4.5$ (lower panel), using 3rd order polynomials in $1/L$. Error bars denote standard errors.}
\label{fig_spin_corr}
\end{figure}
In addition to the above structure factor, we also probed directly
the spin-spin correlation functions
\begin{equation}
C^{a,b}_{s}(\mathbf{x},\mathbf{y}) = \llangle \mathbf{S}_{\mathbf{x}a}\cdot \mathbf{S}_{\mathbf{y}b} \rrangle
\end{equation}
at the largest available distance $\mathbf{d}_L=([L/2+1]-1) \ \mathbf{a}_1+([L/2+1]-1) \ \mathbf{a}_2$ for different system sizes, and performed  a finite size scaling of both
$C^{A,A}_{s}=C^{A,A}_s(0,\mathbf{d}_L)$ and 
$C^{A,B}_{s}=-C^{A,B}_s(0,\mathbf{d}_L)$.
A comparison of the scaling of these quantities to $S_{AF}$ is shown for both $U/t=4$ and $U/t=4.5$ in 
Suppl. Fig.~\ref{fig_spin_corr}, exhibiting the consistency between these different approaches to quantify the spin correlations in the ground state.

\section{Spin excitation gaps}
The gaps for spin excitations at momentum vector $\mathbf{k}$ are obtained from the imaginary-time displaced spin-spin correlation functions for both the staggered sector,
\begin{equation}
S_{s}(\mathbf{k},\tau)= {\llangle (\mathbf{S}_{\mathbf{k}A}(\tau) - \mathbf{S}_{\mathbf{k}B}(\tau))\cdot (\mathbf{S}_{\mathbf{k}A}(0) - \mathbf{S}_{\mathbf{k}B}(0)) \rrangle },
\end{equation}
as well as the uniform sector,
\begin{equation}
S_{u}(\mathbf{k},\tau)=  {\llangle (\mathbf{S}_{\mathbf{k}A}(\tau) + \mathbf{S}_{\mathbf{k}B}(\tau))\cdot (\mathbf{S}_{\mathbf{k}A}(0) + \mathbf{S}_{\mathbf{k}B}(0)) \rrangle },
\end{equation}
where $\mathbf{S}_{\mathbf{k},a}(\tau)=e^{\tau H}\mathbf{S}_{\mathbf{k},a}e^{-\tau H}$. Similarly as for the single-particle gap,  the spin excitation gaps are obtained 
from 
$S_s(\mathbf{k},\tau) \propto \exp(-\tau\Delta_{s}(\mathbf{k}))$, and
$S_u(\mathbf{k},\tau) \propto \exp(-\tau\Delta_{u}(\mathbf{k}))$ 
at large imaginary time $\tau$.
\begin{figure}[hb]
\centering
\includegraphics[width=8cm]{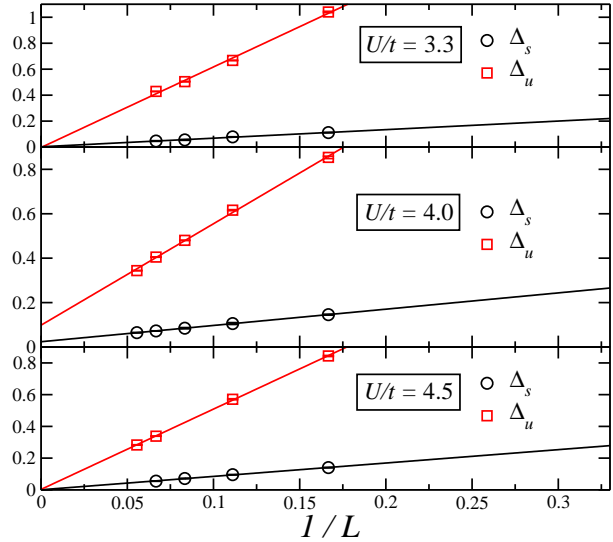}
\caption{Comparison of the finite size scaling between the staggered spin gap $\Delta_{s}$ and the 
uniform spin gap $\Delta_{u}$ at $U/t = 3.3,4$ and $4.5$ (top to bottom). The extrapolated values in the thermodynamic limit for $U/t=4$ are  $\Delta_{s}=0.023\pm0.007$ (s.e.m.) and $\Delta_{u}=0.099\pm0.001$ (s.e.m.). Error bars denote standard errors.}
\label{fig_spin_gaps}
\end{figure}
The staggered spin gap $\Delta_{s}=\Delta_{s}(\Gamma)$ can be calculated  directly via the staggered spin-spin correlations at the $\Gamma$ point. However, since the total magnetization  $ \mathbf{S}_{tot}=\mathbf{S}_{\Gamma A}+\mathbf{S}_{\Gamma B}$ commutes with the Hamiltonian of the system, $[\mathbf{S}_{tot},H]=0$, the uniform spin gap $\Delta_{u}$ cannot be extracted from the uniform spin-spin correlations at the $\Gamma$ point in a canonical quantum Monte Carlo simulation. Instead, one obtains $\Delta_{u}=\lim_{\mathbf{k}\rightarrow \Gamma}\Delta_{u}(\mathbf{k})$ from measurements performed at the finite momenta closest to the $\Gamma$ point for each finite system. Supplementary Fig.~\ref{fig_spin_gaps} shows the finite size data for these gaps at $U/t = 3.3,4$ and $4.5$. For $U/t=4$, both gaps scale to finite values in the thermodynamic limit, with $\Delta_{u}$ being about four times as large as $\Delta_{s}$. For the other two values of $U/t$, both gaps clearly vanish in the thermodynamic limit.

\section{Density correlations}
The density-density correlation function is given by
\begin{equation}
C^{a,b}_{d}(\mathbf{x},\mathbf{y}) = \llangle n_{\mathbf{x}a} n_{\mathbf{y}b} \rrangle,
\end{equation}
where $a,b\in\{A,B\}$. 
At half-filling, $\langle n_{\mathbf{x}a}\rangle=\langle n_{\mathbf{x}b}\rangle=1$. Supplementary Fig. \ref{fig_density_corr} shows the finite size scaling of the density correlations at the largest distance, $C^{A,A}_{d} = |C^{A,A}_{d}(0,\mathbf{d}_L)|$ and $C^{A,B}_{d} = |C^{A,B}_{d}(0,\mathbf{d}_L)|$ at $U/t=4$. Both scale to zero in the thermodynamic limit, and no long-range density correlations persist. Furthermore, in comparison with the spin correlations, the density correlations are seen to be significantly weaker and essentially zero within the statistical error for system sizes $L \ge 9$.  Consistently, we also find no long-range density ordering when analyzing the density structure factor (not shown).
\begin{figure}[b!]
\centering
\includegraphics[width=8cm]{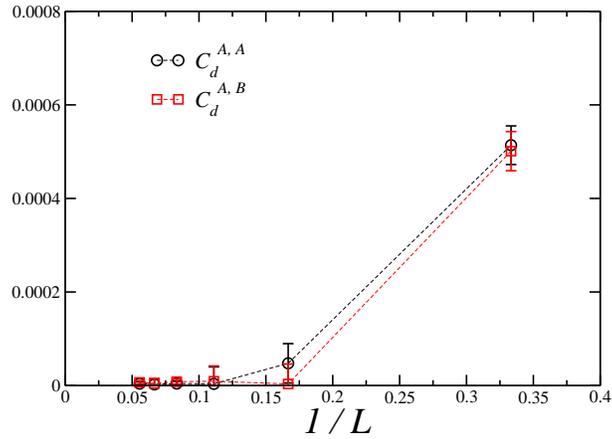}
\caption{Finite size scaling of the density correlation functions $C^{A,A}_{d}$ and $C^{A,B}_{d}$ at $U/t = 4$. Error bars denote standard errors.}
\label{fig_density_corr}
\end{figure}

\section{Dimer-dimer correlations - charge sector}
In this section, we present our results on the dimer-dimer correlations in the charge sector.  We measured both the correlations between the kinetic energy bond operators and the current operators. The spin sector is treated in the following section. 

Correlations between the kinetic energy bond operators
\begin{equation}
k(\mathbf{x},a;\mathbf{y},b)
=\sum_{\alpha}(c^{\dagger}_{\mathbf{x}a\alpha}c^{}_{\mathbf{y}b\alpha}+c^{\dagger}_{\mathbf{y}b\alpha}c^{}_{\mathbf{x}a\alpha}),
\end{equation}
and the current operators 
\begin{equation}
j(\mathbf{x},a;\mathbf{y},b)= 
-i\sum_{\alpha}(c^{\dagger}_{\mathbf{x}a\alpha}c^{}_{\mathbf{y}b\alpha}-c^{\dagger}_{\mathbf{y}b\alpha}c^{}_{\mathbf{x}a\alpha})
\end{equation} 
can be defined between both nearest-neighbour and next-nearest neighbour sites on the honeycomb lattice.
\begin{figure}[hb]
\centering
\includegraphics[width=10cm]{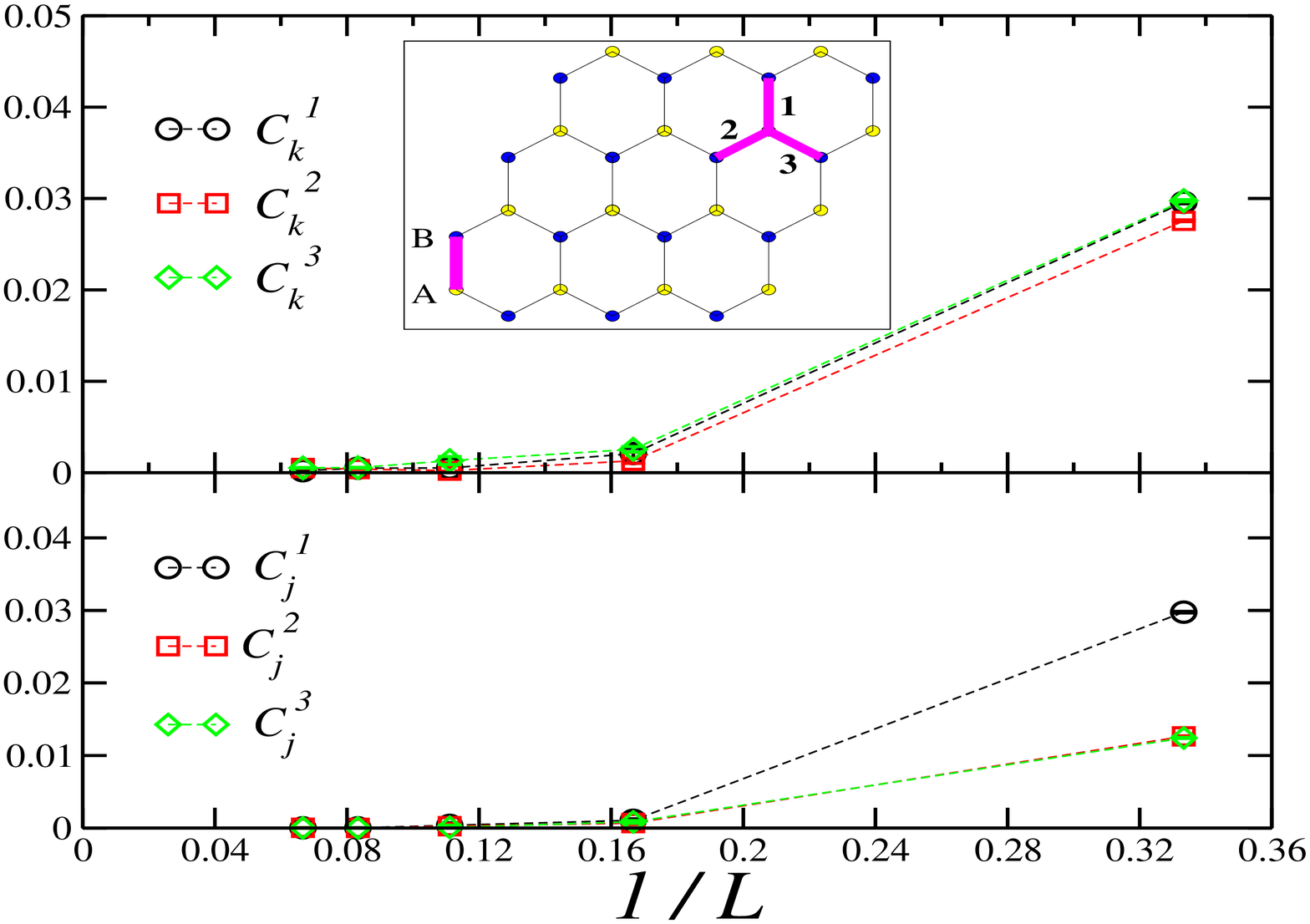}
\caption{Finite size scaling of the nearest-neighbour  correlations $C^{i}_{k}$ and $C^{i}_{j}$ at $U/t=4$. The inset illustrates the three inequivalent directions with respect to the reference bond marked by $AB$. Error bars denote standard errors.}
\label{fig_kinetic_current_corr}
\end{figure}

To probe for VBC order in the kinetic energy sector, we measured the three inequivalent dimer-dimer correlation functions  
\begin{equation}
C^{i}_{k}=|\llangle k(0,A;0,B) k(\mathbf{d}_L,A;\mathbf{d}_L+{\boldsymbol\delta}_i,B)\rrangle|, \quad i=1,2,3,
\end{equation}
at the largest  distance $\mathbf{d}_{L}$ on the finite lattices. For an illustration of the different relative bond 
orientations, see the inset of Suppl.\ Fig.~\ref{fig_kinetic_current_corr}. The upper panel of Suppl.\ Fig.~\ref{fig_kinetic_current_corr} shows the finite size scaling of the $C^{i}_{k}$ at $U/t=4$. These correlations scale to zero in the thermodynamic limit, hence no long-ranged bond order in the kinetic energy persists. Furthermore, in comparison with the spin correlations, these correlations are also seen to be significantly weaker.

To probe for the persistence of nearest-neighbour currents in the ground state, we measured the current-current correlation functions between the bonds of the honeycomb lattice
\begin{equation}
C^{i}_{j}=|\llangle j(0,A;0,B) j(\mathbf{d}_L,A;\mathbf{d}_L+{\boldsymbol\delta}_i,B)\rrangle|, \quad i=1,2,3,
\end{equation}
at the largest  distance $\mathbf{d}_{L}$ on the finite lattices. The corresponding finite size scalings are shown in the lower panel of Suppl. Fig.~\ref{fig_kinetic_current_corr}. Again, long range correlations in the thermodynamic limit can be clearly excluded, indicating the absence of currents between nearest neighbour sites in the ground state. 

\begin{figure}[hb]
\centering
\includegraphics[width=10cm]{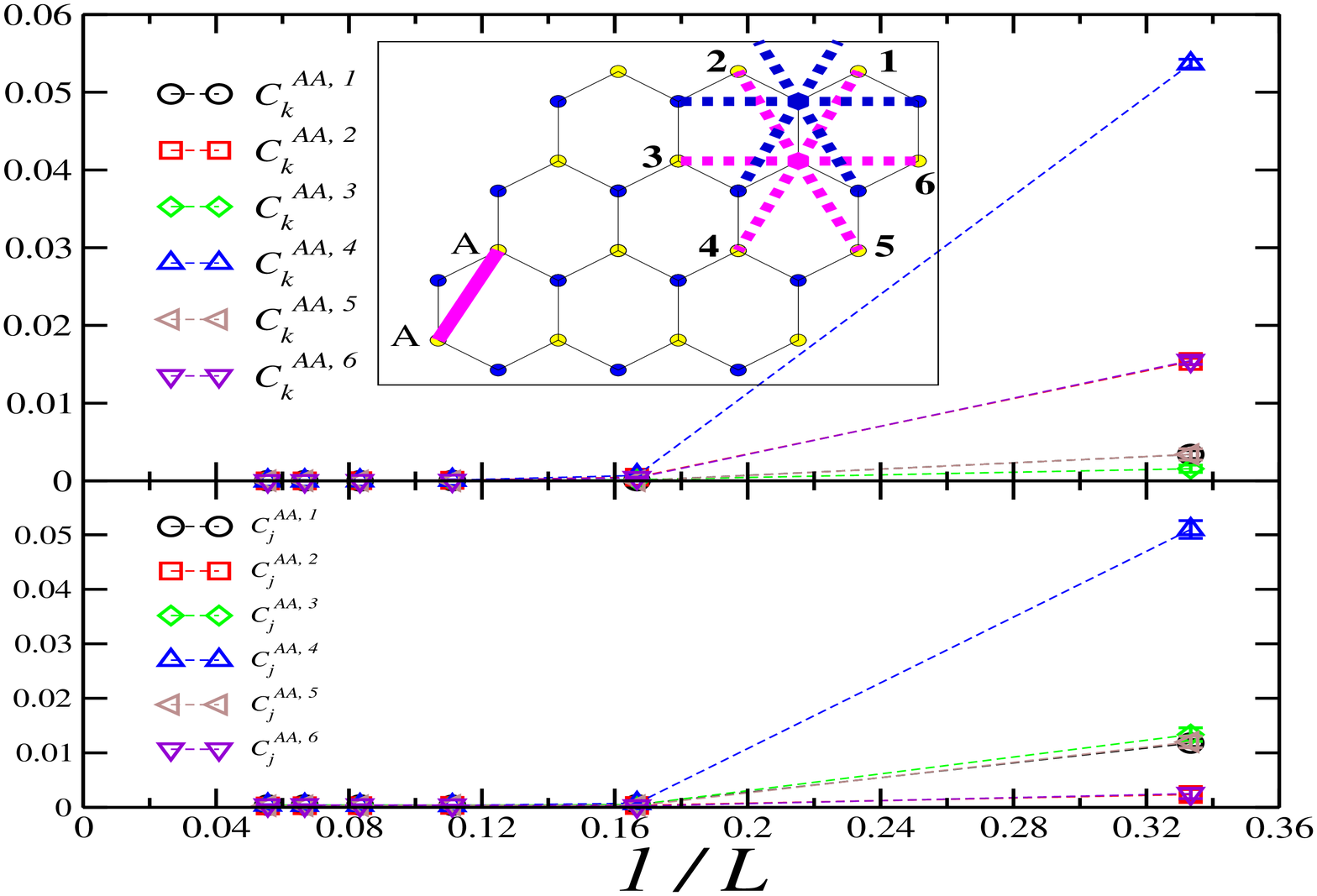}
\caption{Finite size scaling of the  next-nearest neighbour correlations $C^{AA, i}_{k}$ and $C^{AA, i}_{j}$ at the largest distance at $U/t = 4$. The inset illustrates the inequivalent directions with respect to the reference bond marked by $AA$,
with the lower (upper) star corresponding to equal (different) sublattices. Error bars denote standard errors.}
\label{fig_nnn_kin_cur_corr}
\end{figure}

To probe for bond order and currents between next-nearest neighbour sites, we measured
all inequivalent next-nearest neighbour bond-bond and current-current correlation functions at the largest distances both within the same sublattice and  between the two sublattices,
\begin{eqnarray}
C^{AA,i}_{k} &=& |\llangle k(0,A;\mathbf{r}_1,A) k(\mathbf{d}_L,A;\mathbf{d}_L+\mathbf{r}_i,A)\rrangle|, \\
C^{AA,i}_{j} &=& |\llangle j(0,A;\mathbf{r}_1,A) j(\mathbf{d}_L,A;\mathbf{d}_L+\mathbf{r}_i,A)\rrangle|,\\
C^{AB,i}_{k} &=& |\llangle k(0,A;\mathbf{r}_1,A) k(\mathbf{d}_L,B;\mathbf{d}_L+\mathbf{r}_i,B)\rrangle|, \\
C^{AB,i}_{j} &=& |\llangle j(0,A;\mathbf{r}_1,A) j(\mathbf{d}_L,B;\mathbf{d}_L+\mathbf{r}_i,B)\rrangle|,\quad i=1,...,6.
\end{eqnarray}
For an illustration of the different relative bond orientations, see the inset of Suppl.\ Fig.~\ref{fig_nnn_kin_cur_corr}.
The quantum Monte Carlo data for the correlations within the same sublattice at $U/t=4$ 
are shown in Suppl.\ Fig.~\ref{fig_nnn_kin_cur_corr}.
Both $C^{AA,i}_{k}$ and $C^{AA,i}_{j}$ all scale to zero in the thermodynamic limit.  The corresponding correlations between the two sublattices similarly decay to zero in the thermodynamic limit (not shown). Thus no bond ordering  nor currents persist between next-nearest neighbour sites in the ground state at $U/t=4$. 

\section{Dimer-dimer correlations - spin sector}
In the spin sector, we measured the dimer-dimer correlation functions
\begin{equation}
 D_{ij,kl}=\llangle (\mathbf{S}_i\cdot\mathbf{S}_j-\frac{1}{4})(\mathbf{S}_k\cdot\mathbf{S}_l-\frac{1}{4}) \rrangle 
\end{equation}
where $ij$ and $kl$ are each nearest neighbour sites on the honeycomb lattice. The quantum Monte Carlo results for these correlations are shown and discussed in the main text.

\begin{figure}[t!]
\centering
\includegraphics[width=10cm]{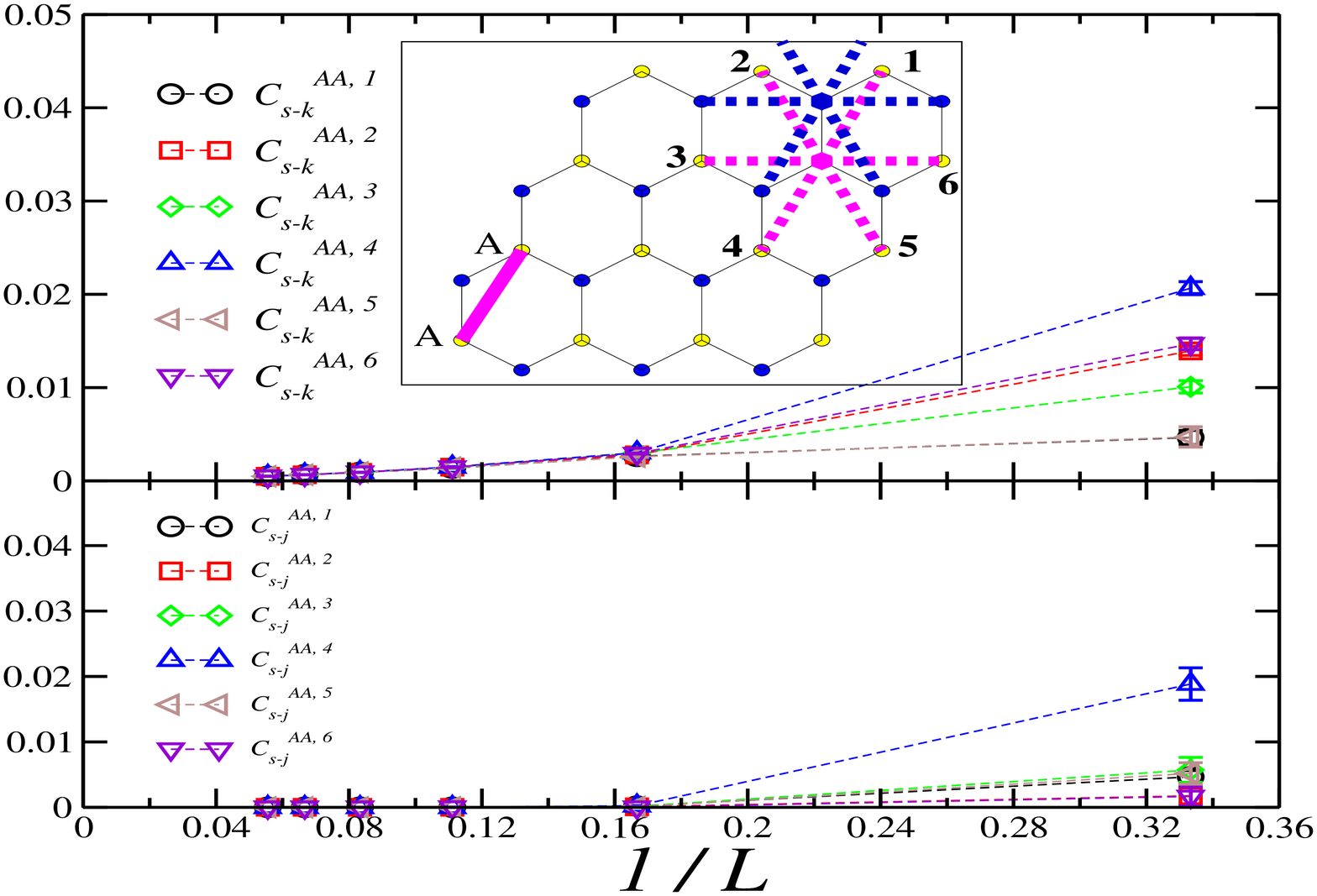}
\caption{Finite size scaling of the next-nearest neighbour correlations $C^{AA, i}_{s-k}$ and $C^{AA, i}_{s-j}$ at the largest distance at $U/t = 4$. The inset illustrates the inequivalent directions with respect to the reference bond marked by $AA$,
with the lower (upper) star corresponding to equal (different) sublattices. Error bars denote standard errors.}
\label{fig_nnn_spinkin_spincur_corr}
\end{figure}

We furthermore measured correlations between the spin-current operators
\begin{equation}
j_s(\mathbf{x},a;\mathbf{y},b)= 
-i\sum_{\alpha}(-1)^\alpha (c^{\dagger}_{\mathbf{x}a\alpha}c^{}_{\mathbf{y}b\alpha}-c^{\dagger}_{\mathbf{y}b\alpha}c^{}_{\mathbf{x}a\alpha})
\end{equation}
as well as the spin-bond operators
\begin{equation}
k_s(\mathbf{x},a;\mathbf{y},b)
=\sum_{\alpha}(-1)^\alpha(c^{\dagger}_{\mathbf{x}a\alpha}c_{\mathbf{y}b\alpha}+c^{\dagger}_{\mathbf{y}b\alpha}c_{\mathbf{x}a\alpha}),
\end{equation}
for next-nearest neighbour sites. We measured these correlations between all inequivalent pairs of next-nearest neighbour sites both within the same sublattice and between the two sublattices at the largest distance on the finite lattices,
\begin{eqnarray}
C^{AA,i}_{s-k}&=&|\llangle k_{s}(0,A;\mathbf{r}_1,A) k_{s}(\mathbf{d}_L,A;\mathbf{d}_L+\mathbf{r}_i,A)\rrangle|,\\
C^{AA,i}_{s-j}&=&|\llangle j_s(0,A;\mathbf{r}_1,A) j_s(\mathbf{d}_L,A;\mathbf{d}_L+\mathbf{r}_i,A)\rrangle|,\\
C^{AB,i}_{s-k}&=&|\llangle k_{s}(0,A;\mathbf{r}_1,A) k_{s}(\mathbf{d}_L,B;\mathbf{d}_L+\mathbf{r}_i,B)\rrangle|,\\
C^{AB,i}_{s-j}&=&|\llangle j_s(0,A;\mathbf{r}_1,A) j_s(\mathbf{d}_L,B;\mathbf{d}_L+\mathbf{r}_i,B)\rrangle|, \quad i=1,..,6.
\end{eqnarray}
For an illustration of the different relative bond orientations, see the inset of Suppl.\ Fig.~\ref{fig_nnn_spinkin_spincur_corr}.
Supplementary Fig.~\ref{fig_nnn_spinkin_spincur_corr} shows the finite size scaling of the correlations within the same sublattice at $U/t=4$.
They all decay to zero in  the thermodynamic limit.
The corresponding correlations between different sublattices
show a similar behavior (not shown).
Thus no spin-bond order nor spin-currents persist between next-nearest neighbour sites in the ground state at $U/t=4$.

\section{Flux quantization measurement for superconductivity}
In the flux quantization measurement, we thread a magnetic flux $\Phi$, in units of the flux quantum $\Phi_{0}$, through the centre of a torus on which the electronic system lies. From the functional form of the ground state energy with respect to the threaded flux, $E_{0}(\Phi/\Phi_{0})$, we can distinguish between normal and superconducting ground states. The signature of the latter requires that the macroscopic energy difference $E_{0}(\Phi/\Phi_{0})-E_{0}(\Phi/\Phi_{0}=1/2)$ scales in the thermodynamic limit
to a periodic function of period $1/2$, and the occurrence of an energy barrier between $\Phi/\Phi_{0}=0$ and $\Phi/\Phi_{0}=1/2$.
In contrast, a metallic 
phase is characterized by ${E_0(\Phi/\Phi_0) - E_0(\Phi/\Phi_0 = 1/2)}$ 
vanishing as a power law 
as a function of system size, while in an insulating phase, it would vanish 
exponentially.
Figure~\ref{fig_flux} compares the QMC 
results 
of the macroscopic energy difference at $U=0$ with that at $U/t=4$.
In both cases one clearly observes the vanishing of this quantity in the thermodynamic limit. Hence,
no signal for superconductivity is obtained from these flux quantization measurements.
\begin{figure}[t!]
\centering
\includegraphics[width=10cm]{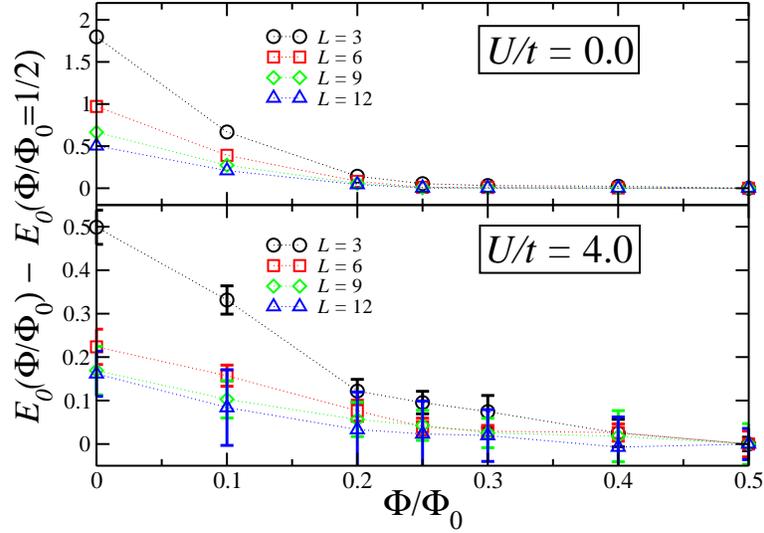}
\caption{The energy difference of $E_{0}(\Phi/\Phi_{0})-E(\Phi/\Phi_{0}=1/2)$ for different system sizes at $U/t=0$ and $U/t=4$. Note that the scale for $U/t=4$ is four times smaller than for $U/t=0$. The flattening of the energy differences exclude the superconducting ground state at both $U/t=0$ and $U/t=4$. Error bars denote standard errors.} 
\label{fig_flux}
\end{figure}

\section{Order parameters for superconductivity}
Order parameters for superconductivity are in principle obtained from considering the irreducible representations of the $D_6$ point group of the honeycomb lattice, which can be described as a triangular lattice with a basis of two atoms in the sublattices $A$ and $B$, respectively. The Cooper pair wave function of a superconducting state 
is a product of a spin, orbital and a sublattice component. Since Pauli's principle requires the wave function to be antisymmetric under particle exchange, we obtain the following possibilities for spin-singlet pairing: for an even (odd) orbital part, the wave function must be symmetric (antisymmetric) under sublattice exchange. 
It is convenient to introduce
pair creation operators in the singlet channel
\begin{equation}
\Delta^\dagger(\mathbf{x},a;\mathbf{y},b)=c^{\dagger}_{\mathbf{x}a\uparrow}c^{\dagger}_{\mathbf{y}b\downarrow}-c^{\dagger}_{\mathbf{x}a\downarrow}c^{\dagger}_{\mathbf{y}b\uparrow},
\end{equation}
where $a,b\in\{A,B\}$.
The operator 
\begin{equation}
\Delta^\dagger_s(\mathbf{x})=\frac{1}{2} ( \Delta^\dagger(\mathbf{x},A;\mathbf{x},A)+\Delta^\dagger(\mathbf{x},B;\mathbf{x},B) )
\end{equation}
describes on-site $s$-wave pairing, which is symmetric under sublattice exchange.  
In Suppl.\ Fig.~\ref{fig_local_pairing}, we show the $s$-wave pair-pair correlation function 
$C_{\Delta}=|\langle \Delta_s(0) \Delta^{\dagger}_s(\mathbf{d}_L)\rangle|$ at the largest distance at $U/t=4$. No long-ranged pairing correlation sustains to the thermodynamic limit; instead, the on-site pair-pair correlation function decreases rapidly.
\begin{figure}[t!]
\centering
\includegraphics[width=10cm]{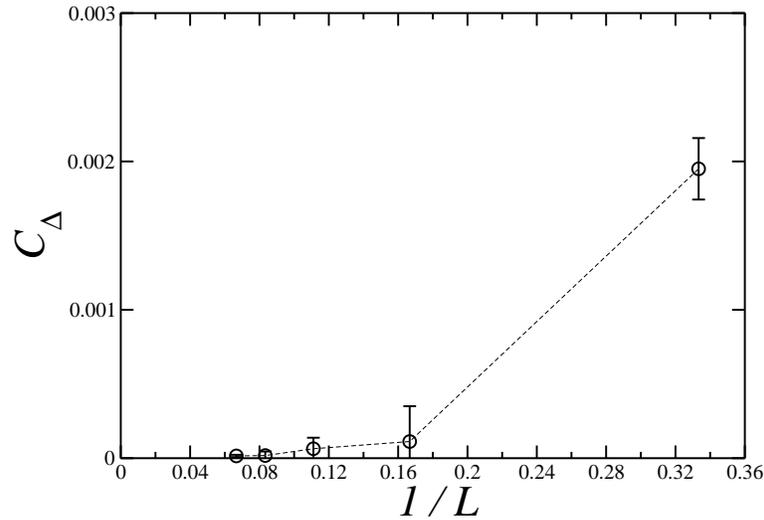}
\caption{Finite size scaling of on-site s-wave pairing correlation $C_{\Delta}$ at $U/t=4$.}
\label{fig_local_pairing}
\end{figure}

Extended pair creation operators based on nearest neighbour pairing can be expressed in terms of phase factors $f^a_1,f^a_2,f^a_3$,
$a\in\{A,B\}$,
\begin{equation}
\Delta^\dagger(\mathbf{x},f^A_1,f^A_2,f^A_3,f^B_1,f^B_2,f^B_3)
=\sum_{i=1}^3 [ f^A_i \Delta^\dagger(\mathbf{x},A;\mathbf{x}+{\boldsymbol\delta}_i,B) + f^B_i \Delta^\dagger(\mathbf{x},B;\mathbf{x}-{\boldsymbol\delta}_i,A) ]
\end{equation}
For an extended $s$-wave, 
\begin{equation}
 \Delta^\dagger_{ext.-s}(\mathbf{x})=\Delta^\dagger(\mathbf{x},1,1,1,1,1,1).
\end{equation}
Nearest neighbour $p$-wave states relate to 
\begin{eqnarray}
\Delta^\dagger_{p_x}(\mathbf{x})&=&\Delta^\dagger(\mathbf{x},0,+1,-1,0,-1,+1),\\
\Delta^\dagger_{p_y}(\mathbf{x})&=&\Delta^\dagger(\mathbf{x},0,+1,+1,0,-1,-1),
\end{eqnarray}
and nearest neighbour $d$-wave states to
\begin{eqnarray}
\Delta^\dagger_{d_{xy}}(\mathbf{x})&=&\Delta^\dagger(\mathbf{x},0,+1,-1,0,+1,-1),\\
\Delta^\dagger_{d_{x^2-y^2}}(\mathbf{x})&=&\Delta^\dagger(\mathbf{x},-2,+1,+1,-2,+1,+1).
\end{eqnarray}
In terms of next-nearest neighbours, one furthermore obtains the singlet $f$-wave state
\begin{equation}
 \Delta^\dagger_{f}(\mathbf{x})=\sum_{j=i}^6 (-1)^i [\Delta^\dagger(\mathbf{x},A;\mathbf{x}+\mathbf{r}_i,A)-\Delta^\dagger(\mathbf{x},B;\mathbf{x}+\mathbf{r}_i,B)].
\end{equation}
\begin{figure}[t!]
\centering
\includegraphics[width=10cm]{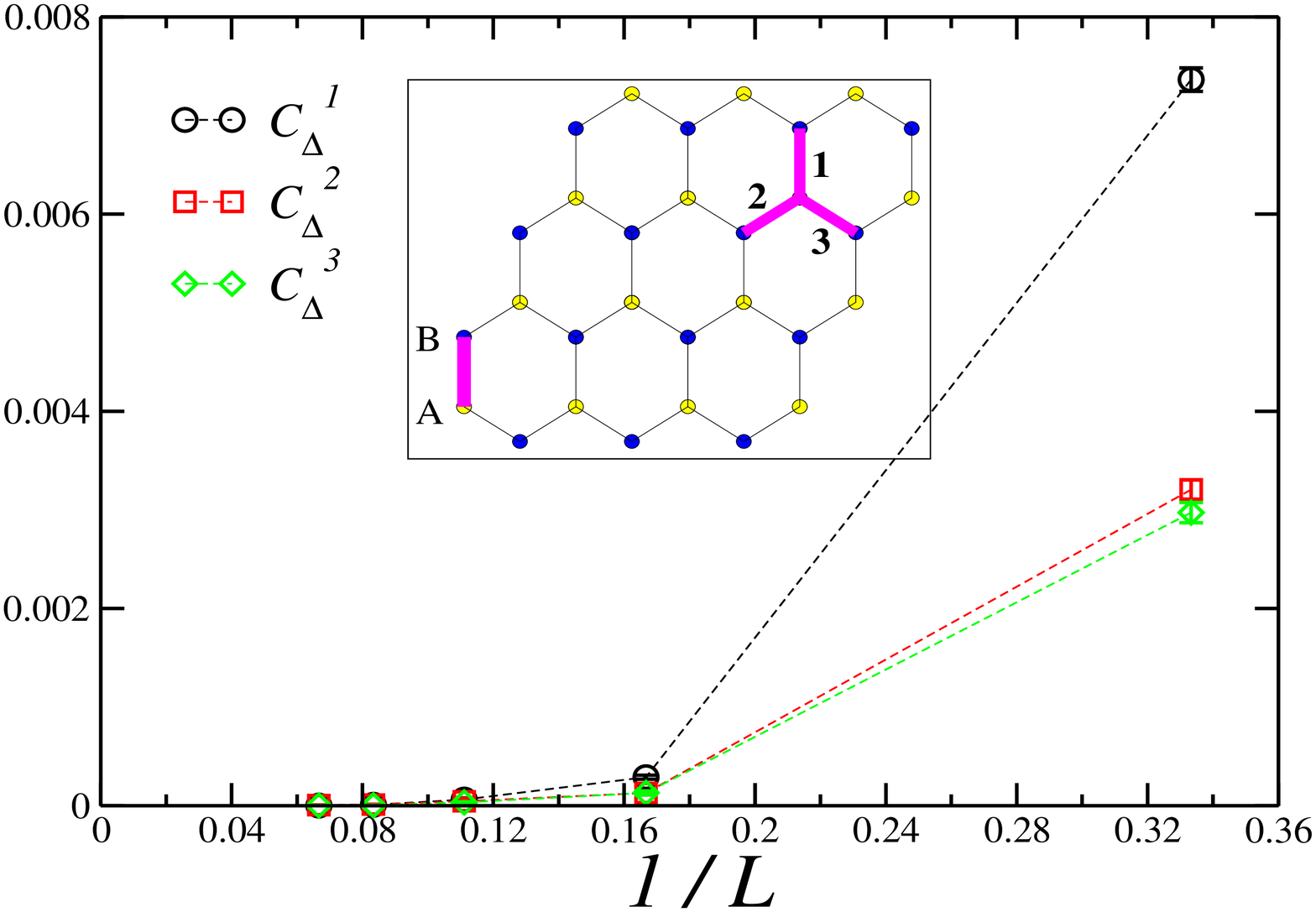}
\caption{Finite size scaling of nearest neigbour pair correlation $C^{i}_{\Delta}$ at $U/t=4$. The inset illustrates the three inequivalent directions with respect to the reference bond marked by $AB$. Error bars denote standard errors.}
\label{fig_nn_pair_corr}
\end{figure}
\begin{figure}[t!]
\centering
\includegraphics[width=10cm]{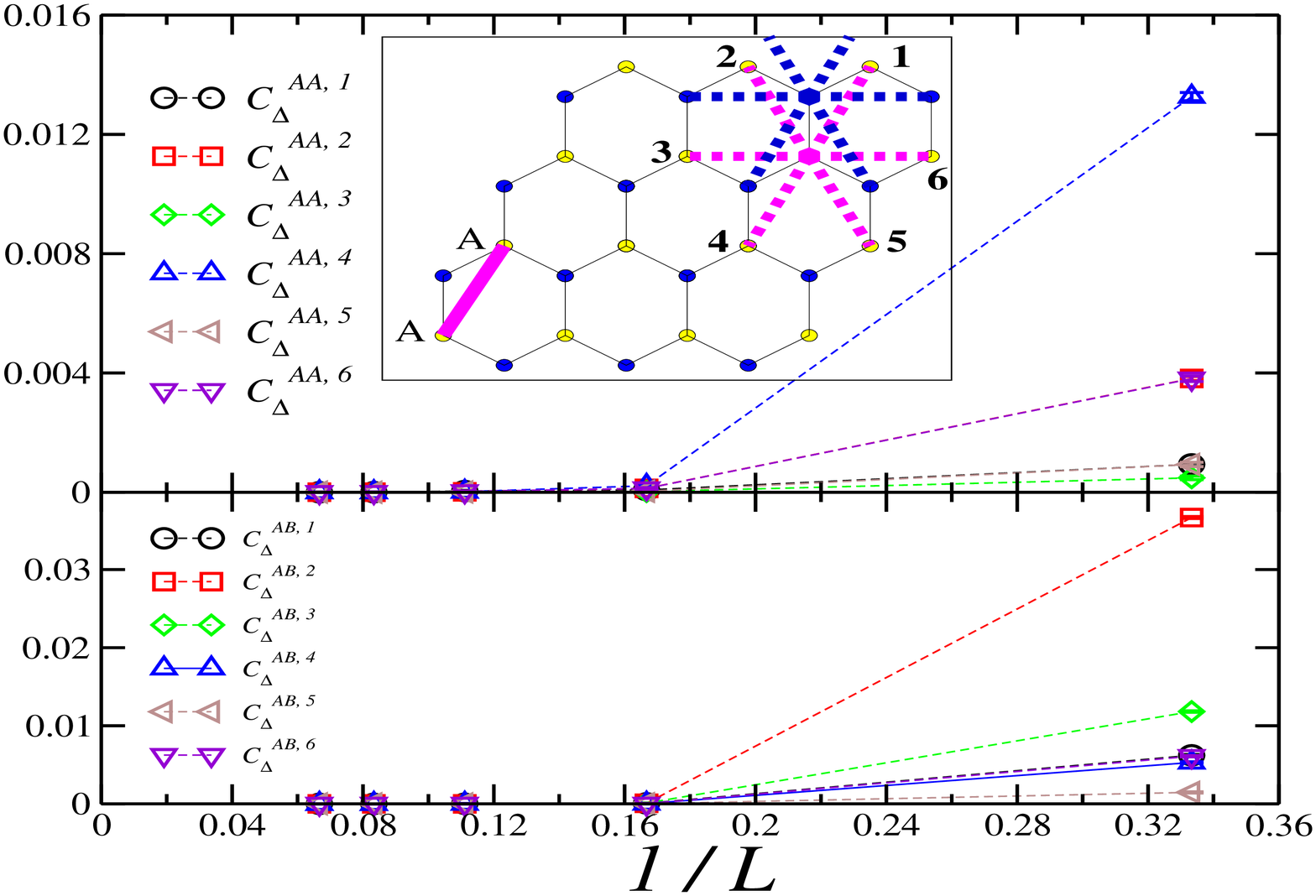}
\caption{Finite size scaling of the next-nearest neighbour pair correlations  $C^{AA, j}_{\Delta}$ and $C^{AB,j}_{\Delta}$ at $U/t=4$.
The inset illustrates the inequivalent directions with respect to the reference bond marked by $AA$,
with the lower (upper) star corresponding to equal (different) sublattices. Error bars denote standard errors.
}
\label{fig_nnn_pair_corr}
\end{figure}

In order to probe for superconductivity based on  nearest neighbour or next-nearest neighbour pairing in the above symmetry sectors, we directly measured in real-space the various inequivalent pair-pair correlation functions at the largest distances on the finite lattices, i.e. 
\begin{equation}
C^i_{\Delta}=|\llangle \Delta(0,A;0,B) \Delta^\dagger(\mathbf{d}_L,A;\mathbf{d}_L+{\boldsymbol\delta}_i,B)\rrangle|,\quad i=1,2,3,
\end{equation}
for the nearest neighbour pairing states, and
\begin{eqnarray}
C^{AA,i}_{\Delta}&=&|\llangle \Delta(0,A;\mathbf{r}_1,A) \Delta^\dagger(\mathbf{d}_L,A;\mathbf{d}_L+\mathbf{r}_i,A)\rrangle| \\
C^{AB,i}_{\Delta}&=&|\llangle \Delta(0,A;\mathbf{r}_1,A) \Delta^\dagger(\mathbf{d}_L,B;\mathbf{d}_L+\mathbf{r}_i,B)\rrangle|,\quad i=1,...,6,
\end{eqnarray}
for next-nearest neighbour pairing states both within the same sublattice and between the two sublattices. As shown in Suppl.\ Fig.~\ref{fig_nn_pair_corr} and Suppl.\ Fig.~\ref{fig_nnn_pair_corr}, both nearest neighbour and next-nearest neighbour pair-pair correlation functions are very weak, even reaching zero within statistical errors for $L \ge 9$. 
From this, we exclude pairing in all the above symmetry sectors, since the full Cooper pair correlations $\langle \Delta_{ext.-s}(0)\Delta^{\dagger}_{ext.-s}(\mathbf{d}_{L})\rangle$, $\langle \Delta_{p_{x}}(0)\Delta^{\dagger}_{p_{x}}(\mathbf{d}_{L})\rangle$, $\langle \Delta_{p_{y}}(0)\Delta^{\dagger}_{p_{y}}(\mathbf{d}_{L})\rangle$, $\langle \Delta_{d_{xy}}(0)\Delta^{\dagger}_{d_{xy}}(\mathbf{d}_{L})\rangle$,
$\langle \Delta_{d_{x^2-y^2}}(0)\Delta^{\dagger}_{d_{x^2-y^2}}(\mathbf{d}_{L})\rangle$, and
$\langle \Delta_{f}(0)\Delta^{\dagger}_{f}(\mathbf{d}_{L})\rangle$
are linear superpositions of the above pair-pair correlation functions, and hence  vanish in the thermodynamic limit. We can thus exclude superconductivity in the half-filled Hubbard model on the honeycomb lattice.

\section{Lower bound for singlet states in the RVB phase}
We consider the internal energy as a function of the projection parameter
$\Theta$,
\begin{equation}
\label{EofTheta1}
E (\Theta) =  
\frac{\langle \Psi_T | {\rm e}^{-\Theta H/2} H {\rm e}^{-\Theta H/2} | \Psi_T \rangle}
{\langle \Psi_T | {\rm e}^{-\Theta H} | \Psi_T \rangle} \; ,
\end{equation}
where $|\Psi_T \rangle$ is the trial wave function (cf. Methods).
Let $\{|n\rangle\}$ be the set of eigenstates of $H$. Then, we have
\begin{equation}
E (\Theta) - E_0 = \frac{1}{1+R(\Theta)}
\sum_{n>0} {\rm e}^{-(E_n-E_0) \Theta} (E_n - E_0) 
\frac{|\langle n|\Psi_T \rangle|^2}{|\langle 0|\Psi_T \rangle|^2} 
\; ,
\end{equation}
where $E_0$ is the ground-state energy, $|0\rangle$ the ground-state, and define
\begin{equation}
R (\Theta) = \sum_{n>0} {\rm e}^{-(E_n-E_0) \Theta} 
\frac{|\langle n|\Psi_T \rangle|^2}{|\langle 0|\Psi_T \rangle|^2} \; .
\end{equation}
Next, we consider  an energy scale
\begin{equation}
\label{DefEpsilon}
\epsilon \ll (E_1 - E_0)/N \; ,
\end{equation}
where $E_1$ is the energy of the first excited singlet with the same quantum numbers as the ground-state,  and define $\Theta^*$, such that for 
$\Theta > \Theta^*$, $(E (\Theta) - E_0)/N < \epsilon$. Typically, $\epsilon$ can be taken of the order of the statistical error in the energy, such that the condition on $\epsilon$ is 
clearly fulfilled. However, $\epsilon$ is not defined in terms of  the statistical errors;  the only defining condition on $\epsilon$ is (\ref{DefEpsilon}).
With such a definition we have
\begin{equation}
\epsilon = \frac{1}{N}\sum_{n>0} \frac{{\rm e}^{-(E_n-E_0) \Theta^*}}{1+R(\Theta^*)} (E_n - E_0) 
\frac{|\langle n|\Psi_T \rangle|^2}{|\langle 0|\Psi_T \rangle|^2} 
\ll (E_1 - E_0)/N
\; ,
\end{equation}
such that 
\begin{equation}
\frac{R(\Theta^*)}{1+R(\Theta^*)} <  \sum_{n>0} \frac{{\rm e}^{-(E_n-E_0) \Theta^*}}{1+R(\Theta^*)} \frac{(E_n - E_0)}{(E_1 - E_0)}
\frac{|\langle n|\psi_T \rangle|^2}{|\langle 0|\psi_T \rangle|^2} 
\ll 1
\; ,
\end{equation}
so that it also holds that $R(\Theta^*) \ll 1$. The last inequality also implies that
\begin{equation}
\label{InequalityForE1}
{\rm e}^{-(E_1-E_0) \Theta^*} \, 
\frac{|\langle 1|\Psi_T \rangle|^2}{|\langle 0|\Psi_T \rangle|^2} \ll 1 \; ,
\end{equation}
since the sum in $R(\Theta)$ consists of positive definite terms.
In  case the overlaps in the last inequality are finite, 
\begin{equation}
(E_1-E_0) \Theta^* \gg 1 \; ,
\end{equation}
such that $1/\Theta^*$ provides a lower bound for $E_1-E_0$.
In  case $ |\langle 1|\Psi_T \rangle|^2/|\langle 0|\Psi_T \rangle|^2 \ll 1$ such that the inequality 
(\ref{InequalityForE1}) is fulfilled due to a vanishing overlap, we miss the lowest excited singlet state, and $1/\Theta^*$ provides a lower bound for the next lowest singlet with a finite overlap with the trial wave function. 
\begin{figure}
\centering
\includegraphics[width=8cm]{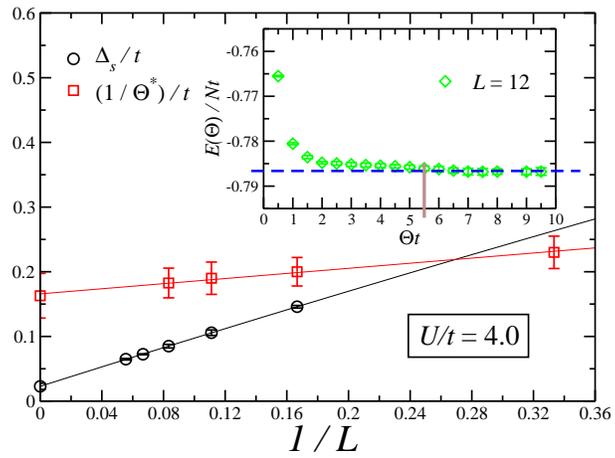}
\caption{Lower bound $1/\Theta^*$ for $L=3$, 6, 9, and 12 at $U/t=4$. For comparison, the values of the 
spin-gap are reproduced. The inset shows the internal energy $E(\Theta)$ as a function of the projection 
parameter $\Theta$ for the $L=12$ system. The vertical bar gives the position of $\Theta^*$. Error bars 
in $\Delta_s$ and $E(\Theta$) denote standard errors.
For the determination of the error bars in $1/\Theta^*$, see the text.}
\label{fig_lower_bound}
\end{figure}

For the determination of the lower bound for singlet states we concentrated on the value
$U/t = 4$, centered in the RVB phase. 
We verified that in the case $L=2$, where the system can be fully diagonalized, 
setting $\epsilon = 10^{-3}t$, $1/\Theta^* = (0.74 \pm 0.04)t$ (s.e.m.) is a lower bound. 
Namely, for this system size, 
$E_1 - E_0= 1.84 t$ for the first excited singlet state of momentum $\mathbf{k} = 0$. 
We also verified that this state has an overlap of 0.22 with the trial wave function. 
The above value of $\epsilon$ corresponds to the maximal error for all system sizes. 
The uncertainty in the determination of $\Theta^*$ is taken as the maximum between 
(i) the distance from $\Theta^*$ to the value of $\Theta$ for $E(\Theta)/N = E(\Theta^*)/N - \epsilon$ and (ii) the distance between two consecutive values of $\Theta$ around $\Theta^*$. By means of error propagation, we then estimate the error in $1/\Theta^*$.
In Suppl.\ Fig.\ \ref{fig_lower_bound} we display $1/\Theta^*$ for $L=3$, 6, 9, and 12, and an extrapolation to the TDL. In all these cases, the lower bound is above the spin-gap, as well as the extrapolation to the TDL. Hence, we find no evidence for singlet states with the same quantum numbers as the ground-state, that may become degenerate with it in the thermodynamic limit. However, this result does no exclude the possibility of low-lying singlet states that have vanishing overlaps with our trial wave function.

\end{SI}

\end{document}